\documentclass[pra,showpacs]{revtex4}
\usepackage[dvips]{graphicx}
\usepackage{amsmath}
\usepackage{graphicx}
\usepackage{makeidx}
\usepackage{subfigure}
\begin{document}
\title{Creation and Reproduction of Model Cells with Semipermeable Membrane}
\author{Hidetsugu Sakaguchi}
\affiliation{Department of Applied Science for Electronics and Materials,\\
Interdisciplinary Graduate School of Engineering Sciences,\\
Kyushu University, Kasuga, Fukuoka 816-8580, Japan}
\begin{abstract}
A high activity of reactions can be confined in a model cell with a semipermeable membrane in the Schl\"ogl model. It is interpreted as a model of primitive metabolism in a cell.  We study two generalized models to understand the creation of primitive cell systems conceptually from the view point of the nonlinear-nonequilibrium physics.  In the first model, a single-cell system with a highly active state confined by a semipermeable membrane is spontaneously created from an inactive homogeneous state by a stochastic jump process. In the second model, many cell structures are reproduced from a single cell, and a multicellular system is created.
\end{abstract}
\maketitle
\section{Introduction}
  Reaction-diffusion systems are important models for various nonlinear processes in biological systems.\cite{rf:1,rf:2,rf:3} Most reaction-diffusion systems have been studied in  homogeneous media to simplify the processes and their analyses. Recently, reaction-diffusion systems in more complicated media such as membranes and microemulsions have been investigated.\cite{rf:4,rf:5} 
 
Complicated biochemical reactions take place in a cell.  
Several researchers considered that chemical reactions in a confined cell-like structure are an important step to life. Oparin considered that  "coacervate"  played an important role in the origin of a cell in prebiotic chemical evolution.\cite{rf:6} He pointed out the importance of chemical reactions confined in the cellular structure "coacervate". Dyson pointed out the importance of the jump to a highly activated state in a certain bistable system as the origin of life.\cite{rf:7} Recently, Noireaux and Libchaber have studied  a cell-like bioreactor using a phospholipid vesicle.\cite{rf:8}  

A cell is separated from the outside world by a membrane. A cell becomes an independent element owing to the existence of the membrane. 
Complicated chemical reactions occur only inside of the cell.  Some materials and reactants are transported through the membrane.  
The consumption of nutrients and elimination of waste matter through the membrane is a basic process of metabolism. A highly active nonequilibrium state is maintained only inside of the cell.  
Inside and outside of the cell, materials are in an aqueous solution, but the membrane is composed of lipid. Therefore, the transport coefficient in the membrane is different from that inside and outside of the cell.  
We constructed a simple reaction-diffusion system based on the Schl\"ogl model to understand the confined nonequilibrium states accompanying the transport of materials through the membrane qualitatively.\cite{rf:9}  In the model, the Schl\"ogl reaction proceeds in a vesicle with a semipermeable membrane, where the diffusion constant of some materials is small or zero. The semipermeable membrane is necessary to confine the activity inside of the cell. 
Because complicated reactions including those of DNA and RNA, and active transport through the membrane were not included in the model, our model is too simple as a model of a living cell. 
Our model is not a realistic model, but an artificial model for considering "metabolism" conceptually from the view point of a dissipative structure far from equilibrium. 
 In this paper, we propose two models for generalizing the previous model. In the first model, a single-cell system with a highly active state confined by a semipermeable membrane is created from an inactive homogeneous state by a stochastic jump.  In the second model, semipermeable membranes are reproduced and a multicellular system is created. 

\section{Schl\"ogl Model}
The Schl\"ogl model is a set of chemical reactions including three kinds of chemicals, namely, U, V and W \cite{rf:10} represented as 
\[
{\rm V}+{\rm 2U}\rightleftharpoons {\rm 3U},\; {\rm U}\rightleftharpoons {\rm W}.\]
We interpret the three chemicals V, U and W as a nutrient material, a self-catalytic product and a waste material, respectively.
The reaction-diffusion equations for $u=[{\rm U}]$, $v=[{\rm V}]$, and $w=[{\rm W}]$ are expressed as 
\begin{eqnarray}
\frac{\partial u}{\partial t}&=&k_1vu^2-k_2u^3-k_3u+k_4w+D_u\nabla^2 u,\nonumber\\
\frac{\partial v}{\partial t}&=&-k_1vu^2+k_2u^3+D_v\nabla^2 v,\nonumber\\
\frac{\partial w}{\partial t}&=&k_3u-k_4w+D_w\nabla^2w.
\end{eqnarray}
where $k_1$, $k_2$, $k_3$  and $k_4$ are the rate constants for the reactions V$+$2U$\rightarrow$3U, 3U$\rightarrow$V$+$2U, U$\rightarrow$W, and W$\rightarrow$ U, respectively, and $D_u$, $D_v$, and $D_w$ are the diffusion constants of the materials. In general reaction-diffusion equations, the diffusion constants are assumed to be uniform in space. However, we assume a vesicle surrounded by a membrane as a model of a primitive cell. The diffusion constants inside of the vesicle are different from those in the bulk region.  We assume nonuniform diffusion constants as 
\begin{eqnarray}
D_u(r)&=&D_{u1}, \;\;\ {\rm for}\;\; r<r_0,\; r>r_0+d,\nonumber\\
 &=&D_{u0} \;\;\;{\rm for}\;\; r_0\le r \le r_0+d,\nonumber\\
D_v(r)&=&D_{v1}, \;\;\ {\rm for}\;\; r<r_0,\; r>r_0+d,\nonumber\\
 &=&D_{v0} \;\;\;{\rm for}\;\; r_0\le r \le r_0+d,\nonumber\\
D_w(r)&=&D_{w1}, \;\;\ {\rm for}\;\; r<r_0,\; r>r_0+d,\nonumber\\
 &=&D_{w0} \;\;\;{\rm for}\;\; r_0\le r \le r_0+d,
\end{eqnarray}
where $r$ is the distance from the center of a spherical cell. 
The value $r_0+d$ represents the size of the cell, and $d$ is the thickness of the membrane.  The diffusion constants are uniform inside and outside of the cell. 
The diffusion constants inside of the cell might be different from those outside of the cell, but we assume the same values for the sake of simplicity.  In all simulations in this paper, we have assumed $D_{u1}=D_{v1}=D_{w1}=D_1$ and $D_{u0}=D_{v0}=D_{w0}=D_0$.    
The diffusion constants are assumed to be smaller in the membrane region $r_0\le r \le r_0+d$. That is, the transport of materials by diffusion  is assumed to be more difficult in the membrane region. 
In our previous research, we have studied several other cases. In the case of $D_{u0}=0, D_{v0}=D_{w0}\ne 0$, a confined active state was obtained, which is similar to that in the case in this paper.  In another case of $D_{v0}=0,D_{u0}=D_{w0}\ne 0$, no confined active state was obtained because the nutrient cannot be supplied to the cell. 

Figure 1(a) shows an example of a confined active state at $k_1=k_2=k_3=k_4=1,D_1=1, D_0=0.01$, $r_0=2$, and $d=1$. The boundary conditions are assumed to be $u(R)=w(R)=0$ and $v(R)=v_0=0.8$ at a radius $R=20$ located far from the center of the cell. The initial conditions are $v(r)=v_0,w(r)=0$, and $u(r)=0$ for $r>r_0+d$, and $u(r)=0.07$ for $r\le r_0+d$. By assuming a spherical symmetry, $\nabla^2$ is replaced by $d^2/dr^2+(2/r)(d/dr)$ in the numerical simulation.  A high activity of the self-catalytic product U is maintained inside of the cell owing to the low diffusivity of the membrane.\cite{rf:9}  
Figure 1(b) shows another stationary state called the death state: $v(r)=v_0$ and $u(r)=w(r)=0$, because no self-catalytic product is produced and 
the concentrations $u$, $v$ and $w$ take the same values inside and outside of the cell membrane. This state was obtained from the initial conditions: $v(r)=v_0,w(r)=0$, and $u(r)=0$ for $r>r_0+d$ and $u(r)=0.06$ for $r\le r_0+d$. The active and death states are bistable when $D_0=0.01$ and $D_1=1$. When $D_0=D_1=1$, only the death state appears if the other parameters are fixed. 

\begin{figure}[t]
\begin{center}
\includegraphics[height=4.5cm]{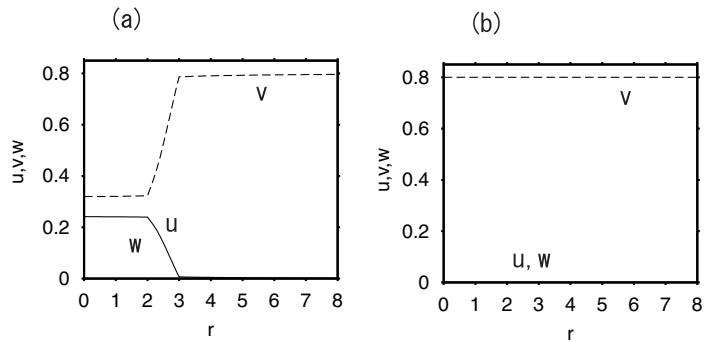}
\end{center}
\caption{(a) Profiles of $u(r),v(r)$ and $w(r)$ for the active state obtained using  eq.~(1) for $k_1=k_2=k_3=k_4=1,D_1=1, D_0=0.01$, $x_0=2$, and $d=1$. The initial conditions are $v(r)=0.8,w(r)=0$, and $u(r)=0$ for $r>r_0+d$ and $u(r)=0.07$ for $r\le r_0+d$. (b) Death state obtained from the initial conditions: $v(r)=0.8,w(r)=0$ and $u(r)=0$ for $r>r_0+d$ and $u(r)=0.06$ for $r\le r_0+d$.}
\label{f1}
\end{figure}
\begin{figure}[t]
\begin{center}
\includegraphics[height=4.5cm]{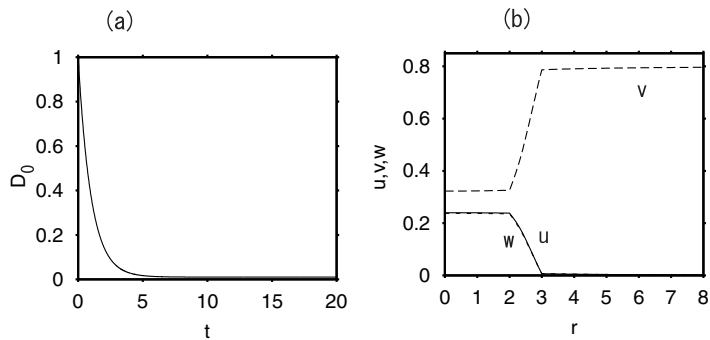}
\end{center}
\caption{(a) Time evolution of the diffusion constant $D_0(t)$ of the membrane. (b) Localized solutions: $u(r)$, $v(r)$, and $w(r)$ to eqs.~(1) and (3) with $u_{th}=0.03$.}
\label{f2}
\end{figure}
In a living cell, a material Z such as lipid, from which the cell membrane is constructed, is synthesized by the cell itself.  We can construct a simple model in which the material generating the low diffusivity of the cell membrane is synthesized via the self-catalytic product U as
\begin{equation}
\frac{dz}{dt}=\frac{c}{4\pi/3(r_0+d)^3}\int_0^{r_0+d}\theta(u-u_{th})4\pi r^2dr-k_zz,
\end{equation}
where $z$ is the concentration $[{\rm Z}]$ of Z, $u_{th}$ is the threshold of $u$ for the creation of the material Z, and $\theta(x)$ is the Heaviside step function. 
Here, we assume that Z is produced when $u$ is larger than $u_{th}$ with a constant rate $c$ and is decomposed with a rate constant $k_z$ just as in a simple model.  The diffusion constant $D_1=1$ and $D_0$ is assumed to linearly decrease as $D_0=1-0.99z$ with the existence of the material Z such as lipid. This is also a simple model. We have performed numerical simulation for $c=1,u_{th}=0.03$, and $k_z=1$ from the initial condition: $z=0$, $D_0=1$, $v(r)=v_0=0.8,w(r)=0$, and $u(r)=0$ for $r>r_0+d$ and $u(r)=1$ for $r\le r_0+d$. That is, we have assumed a high activity of $U$ inside of the cell as an initial condition. 
Figure 2(a) shows the time evolution of $D_0$. As the material Z is produced by a high activity of U inside of the cell, $D_0$ decreases from 1 towards 0.01.
The low diffusivity of the membrane maintains the high activity of U inside of the cell. The low diffusivity of the membrane is maintained by the high density of the self-catalytic product $U$. If $D_0$ is close to 1 for a long time, the concentration of U inside of the cell decreases to 0, which leads to the death state. In the death state, $z=0$ and $D_0$ remains to be 1.  Figure 2(b) shows an active state that appeared as a result of the numerical simulation. The profiles of $u$, $v$ and $w$ are almost the same as those in Fig.~1(a). 
In this model, the low diffusivity of the membrane and the high activity of $U$ inside of the cell are mutually dependent on each other. In other words, an active state confined in the membrane with a low diffusivity does not appear naturally from a death state in a system of uniform diffusivity $D_0=D_1$.

\begin{figure}[t]
\begin{center}
\includegraphics[height=4.5cm]{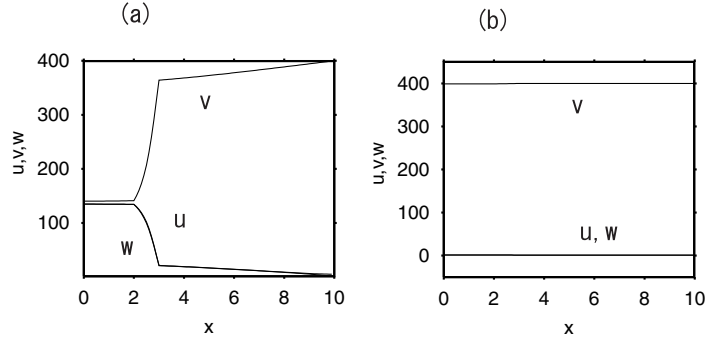}
\end{center}
\caption{Bistable states in a one-dimensional model. (a) Active state. (b) Death state.}
\label{f3}
\end{figure}
\section{Stochastic Schl\"ogl Model}
For the comparison with the stochastic model introduced later, we first consider a one-dimensional model where $\nabla^2$ in eq.~(1) is replaced by $\partial^2/\partial x^2$.  Figures 3(a) and 3(b) show an active state and a death state in the one-dimensional model for  $k_1=k_2=8\times 10^{-6},k_3=k_4=2,D_1=2, D_0=0.02$, $x_0=2$, and $d=1$. The boundary conditions are $\partial u/\partial x=0,v=400$, and $w=1$ at $x=10$, and $\partial u/\partial x=\partial v/\partial x=\partial w/\partial x=0$ at $x=0$. 
The two states are bistable similarly to that in the case of the spherically symmetric model shown in Fig.~1. 
The active state confined by the membrane of low diffusivity is also not naturally created from a death state in this deterministic model. 

We construct a stochastic version of the Schl\"ogl model as a model of the birth of a primitive cell.
Stochastic chemical reactions have been studied as nonlinear-nonequilibrium systems  far from equilibrium.\cite{rf:1}  The stochasticity originates from the finiteness of the number of molecules. A novel state generated by the discreteness of molecules was reported in a small autocatalytic system.\cite{rf:11}
Recently, several stochastic models have been studied to understand fluctuations in gene networks.\cite{rf:12,rf:13}

The one-dimensional space is further simplified to a discrete one-dimensional lattice of interval 1, and the time is also discretized. The numbers of the molecules U, V, and W at the $i$th site and the $n$th step are respectively denoted by $N_u(i,n),N_v(i,n)$, and $N_w(i,n)$  where $i=0,1,2,\cdots,M$. The probability for the reaction V$+$2U$\rightarrow$3U to take place at the $i$th site and  $n$th step is assumed to be $k_1N_v(i,n)N_u^2(i,n)$.  
That is, the numbers of U and V change by 1 as $N_u(i,n+1)=N_u(i,n)+1$ and $N_v(i,n+1)=N_v(i,n)-1$ with the probability $k_1N_v(i,n)N_u^2(i,n)$. The numbers of U and V do not change with the probability $1-k_1N_v(i,n)N_u^2(i,n)$. 
Similarly, the probability for the reaction 3U$\rightarrow$V$+$2U is assumed to $k_2N_u(i,n)^3$, and the probabilities of the reaction ${\rm U}\rightarrow {\rm W}$ and the inverse reaction are respectively  $k_3N_u(i,n)$ and $k_4N_w(i,n)$. 
We perform a Monte-Carlo simulation in which the numbers of molecules, namely $N_u$, $N_v$, and $N_w$ exhibit random walks according to the probabilities determined by the molecule numbers at each site.  The diffusion processes are also simulated by another random walk between neighboring sites. That is, the number $N_u(i)$ of U at the $i$th site increases by 1 as $N_u(i,n+1)=N_u(i,n)+1$, and $N_u(i+1)$ at the $(i+1)$th site decreases by 1 as $N_u(i+1,n+1)=N_u(i+1,n)-1$ with the probability $D_u(i+1/2)N_u(i+1,n)$. Inversely, $N_u(i)$ decreases by 1 as $N_u(i,n+1)=N_u(i,n)-1$ and $N_u(i+1)$ increases by 1 as $N_u(i+1,n+1)=N_u(i+1,n)+1$ with the probability of $D_u(i+1/2)N_u(i,n)$. Each diffusion constant $D_u(i+1/2)$ is defined at the bond between the $i$th site and the $(i+1)$th site. The diffusion constant is assumed to be $D_u(i+1/2)=D_1$ except for $i=2$ and $D_u(i+1/2)=D_0$ only for $i=2$. At the left edge $i=0$, $D_u(i+1/2)$ is assumed to be 0. The reaction and diffusion processes are performed independently in our Monte-Carlo simulation.  

\begin{figure}[t]
\begin{center}
\includegraphics[height=4.cm]{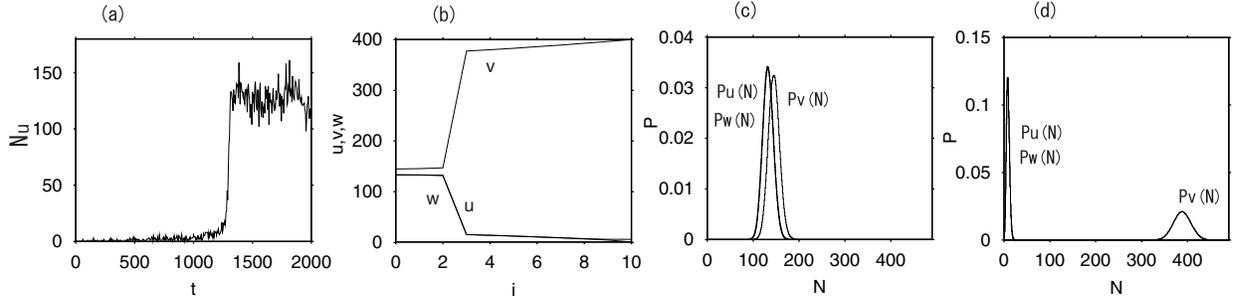}
\end{center}
\caption{(a) Time evolution of $N_u(0)$. (b) Time averages $u,v$, and $w$ of $N_u(i),N_v(i)$, and $N_w(i)$. (c) Probability distributions $P_u(N),P_v(N)$ and $P_w(N)$ of $N_u(i),N_v(i)$ and $N_w(i)$ at $i=1$. (d) Probability distributions $P_u(N),P_v(N)$, and $P_w(N)$ of $N_u(i),N_v(i)$, and $N_w(i)$ at $i=7$.
$P_u(N)$ and $P_w(N)$ almost overlap in (c) and (d) . 
}
\label{f4}
\end{figure}
If the fluctuations of $N_u,N_v$, and $N_w$ are neglected, that is, $N_u,N_v$, and $N_w$  are assumed to be equal to their ensemble averages $\langle N_u\rangle, \langle N_v\rangle $ and $\langle N_w\rangle$,  similarly, $N_u^2N_v$ and $N_u^3$ are assumed  to be equal to $\langle N_u\rangle^2\langle N_v\rangle$ and $\langle N_u\rangle^3$, rate equations  for $N_u,N_v$, and $N_w$ are obtained as 
\begin{eqnarray}
& &\langle N_u(i,n+1)\rangle=\langle N_u(i,n)\rangle+k_1\langle N_v(i,n)\rangle \langle N_u(i,n)\rangle^2-k_2\langle N_u(i,n)\rangle^3-k_3\langle N_u(i,n)\rangle+k_4\langle N_w(i,n)\rangle \nonumber\\
&+&D_u(i+1/2,n)(\langle N_u(i+1,n)\rangle-\langle N_u(i,n)\rangle)+D_u(i-1/2,n)(\langle N_u(i-1,n)\rangle-\langle N_u(i,n)\rangle),\nonumber\\
& &\langle N_v(i,n+1)\rangle=\langle N_v(i,n)\rangle-k_1\langle N_v(i,n)\rangle\langle N_u(i,n)\rangle^2+k_2\langle N_u(i,n)\rangle^3\nonumber\\&+&D_v(i+1/2,n)(\langle N_v(i+1,n)\rangle-\langle N_v(i,n)\rangle)+D_v(i-1/2,n)(\langle N_v(i-1,n)\rangle-\langle N_v(i,n)\rangle),\nonumber\\
& &\langle N_w(i,n+1)\rangle=\langle N_w(i,n)\rangle +k_3\langle N_u(i,n)\rangle-k_4\langle N_w(i,n)\rangle\nonumber\\&+&D_w(i+1/2,n)(\langle N_w(i+1,n)\rangle-\langle N_w(i,n)\rangle)+D_w(i-1/2,n)(\langle N_w(i-1,n)\rangle-\langle N_w(i,n)\rangle).
\end{eqnarray}
If the difference operators in both time and space in eq.~(4) are replaced with a differential operator, the original rate equation eq.~(1) is recovered.

\begin{figure}[t]
\begin{center}
\includegraphics[height=4.5cm]{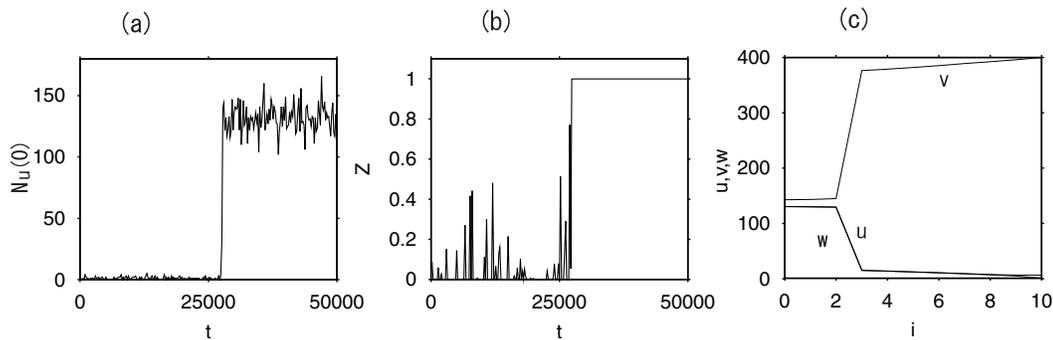}
\end{center}
\caption{(a) Time evolution of $N_u(0)$. (b) Time evolution of $Z$. (c) Long-time averages of $N_u(i),N_v(i)$, and $N_w(i)$.}
\label{f5}
\end{figure}
We have performed a direct Monte-Carlo simulation for $k_1=k_2=8\times10^{-10},k_3=k_4=2\times 10^{-4},D_1=2\times 10^{-4}$, and $D_0=0.02\times 10^{-4}$, and $N_v$ and $N_w$  are respectively fixed to be $N_v=400$ and $N_w=1$ at the boundary site $i=M=10$. 
The relative ratios of these parameters are the same as those in the case of the deterministic one-dimensional model shown in Fig.~3.  
 The initial condition was a death state, that is, $N_u(i)=N_w(i)=0$ and $N_v(i)=400$. Figure 4(a) shows the time evolution of $N_u(0)$ at the left edge site. $N_u(0)$ increases rapidly at $t=1300$. Here, the time $t$ is determined using $t=n\Delta t$, where $\Delta t=10^{-4}$.  
 It is a stochastic jump process from a death state to an active state. We observed no inverse transition from the active state to the death state until the final time of the simulation $t=10000$.
Figure 4(b) shows  long-time averages of $N_u(i),N_v(i)$, and $N_w(i)$ after a transition time. 
The profiles are fairly close to those for the deterministic model shown in Fig.~3, although the space is continuous in the deterministic model. 
Figures 4(c) and (d) show the probability distributions $P_u(N), P_w(N)$, and $P_v(N)$ for $N_u(i), N_w(i)$, and $N_v(i)$ at two sites $i=1$ and $i=7$. The probability distribution of $P_w(N)$ almost overlaps with $P_u(N)$ in these figures. The stochastic transition has occurred owing to the fluctuations in $N_u,N_v$, and $N_w$.  

We can further construct a model in which the material Z producing the low diffusivity of the membrane is synthesized by the self-catalytic product U as
\begin{equation}
Z(n+1)=Z(n)+\Delta t\{\theta(\sum_{i=0}^2 N_u(i,n)-u_{th})-Z(n)\},
\end{equation}
where $\Delta t=10^{-4}$ and $u_{th}$ is the threshold of $u$ for generating the material. The diffusion constant $D_0$ in the membrane region is assumed to be given by $D_0=D_1(1-0.99Z(n))$. The quantity $Z$ fluctuates in time because of the temporal fluctuation in $N_u$. The initial condition is $D_0=D_1=2\times 10^{-4}$, and therefore $Z(0)=0$.   At the boundary site $i=M=10$, $N_v$ and $N_w$  are respectively fixed to be $N_v=400$ and $N_w=1$. The initial condition is a death state $N_u(i)=N_w(i)=0$ and $N_v(i)=400$. (Almost the same result was obtained from another initial condition $N_u(i)=N_w(i)=N_v(i)=0$.)   Figures 5(a) and 5(b) show the time evolution of $N_u(0)$ at the left edge $i=0$ and $Z(n)$ for $u_{th}=8$.   Initially, $N_u(0)$ exhibits fluctuation around 1, and $Z$ increases intermittently from 0, but it decreases to nearly 0 again for $t<27500$. However, $N_u(0)$ increases rapidly to $N_u=130$ at $t\sim 27500$, and then $Z$ increases to 1. After that,  $Z$ is fixed to 1 and then $D_0$ is fixed to $0.01D_1$.    
Figure 5(c) shows  long-time averages of $N_u(i),N_v(i)$, and $N_w(i)$ after a transition time, which is almost equal to those in Fig.~4(b). 
In a uniform system with $D_0=D_1$, only the death state is stable.  Therefore, the semipermeable membrane with low diffusivity and the active state inside of the cell  should be created almost simultaneously. Actually, such an active state surrounded by a semipermeable membrane is spontaneously created from a death state in a system with an almost homogeneous diffusivity by a stochastic jump process at $t\sim 27500$.  This stochastic jump is more difficult than the simple stochastic jump in a bistable system studied in the previous model shown in Fig.~4.  
The high activity inside of the cell is maintained and protected by the semipermeable membrane, and the membrane that determines the form of the cell is created by the materials inside of the cell.  Both are necessary for the construction of a cell. Therefore, the stochastic jump process might be interpreted as the birth of a cell, and the jump process might also be interpreted as one of the sequential jump processes toward the origin of life, as Dyson has pointed out.

\begin{figure}[t]
\begin{center}
\includegraphics[height=4.5cm]{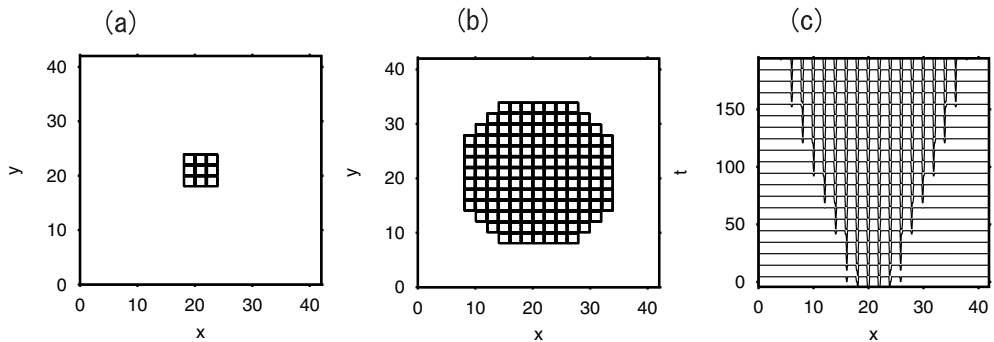}
\end{center}
\caption{Snapshot profiles of $D(x,y)$ at (a) $t=5$ and (b) $t=150$ for $v_0=1.2$. The shaded regions represent the semi-permeable membrane where the diffusion constant $D(x,y)$ is smaller than 0.4. (c) Time evolution of $D(x,y,t)$ at the section of $y=L/2$.}
\label{f6}
\end{figure}
\begin{figure}[t]
\begin{center}
\includegraphics[height=4.5cm]{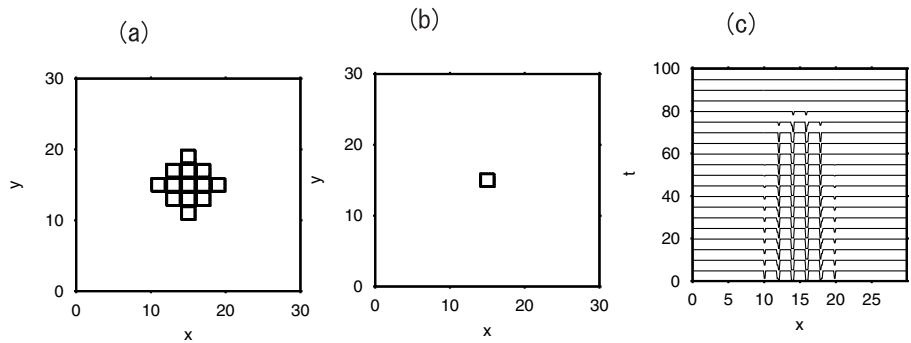}
\end{center}
\caption{Snapshot profile of $D(x,y)$ at (a) $t=5$ and (b) $t=75$ for $v_0=0.7$. The shaded regions represents the semi-permeable membrane where the diffusion constant $D(x,y)$ is smaller than 0.4. (c) Time evolution of $D(x,y,t)$ at the section of $y=L/2$.}
\label{f7}
\end{figure}
\section{Reproduction and Extinction of Model Cells}
In \S 2, we have constructed a single cell model with a spherical symmetry. 
The low diffusivity in the membrane region is maintained by the material Z produced inside of the cell. 
We can construct a two-dimensional model where $\nabla^2$ in eq.~(1) represents $\partial^2/\partial x^2+\partial^2/\partial y^2$. Each cell is assumed to have a square form of size $l=2$, and the width $d$ of the membrane for each cell is assumed to be 0.2. 
The diffusion constant $D_0$ in the membrane region for each cell at $(i,j)$  site is given by  $D_0(i,j)=(1-0.99z(i,j))D_1$. The material Z at the $(i,j)$ site is assumed to be produced by the cell surrounded by the membrane as
\begin{equation}
\frac{dz(i,j)}{dt}=\frac{1}{l^2}\int_{il}^{(i+1)l}\int_{jl}^{(j+1)l}\theta(u(x,y)-u_{th})dxdy-z(i,j).
\end{equation}
We  performed numerical simulations of eqs.~(1) and (6). 
The parameters are $k_1=k_2=k_3=k_4=1$ and $D_1=1$.
The boundary conditions are $v=v_0, u=w=0$ for $r>21$. The nutrient concentration $v_0$ at the boundary is a control parameter. The initial condition is $z(i,j)=0$, $v(x,y)=v_0, w(x,y)=0$, and $u(x,y)=1$ for $r<l_0$ and $u(x,y)=0$ for $r>l_0$, where $r=\sqrt{(x-L/2)^2+(y-L/2)^2}<2$ is the distance from the center of the system of size $L$.  That is, only the central region of size $l_0$ is activated as an initial condition. The membrane is created at the central site by the existence of U. If the membrane is constructed around the cell region, the concentration of the self-catalytic product is maintained inside of the membrane, and a single cell is created at the central site.  By the slow diffusion of U through the membrane, U increases in neighboring sites, leading to the construction of the membrane in  neighboring sites, and new cells are created around the central cell. 
In this deterministic model, an active cell cannot be created from a death state as in the previous section, but a multicellular structure can appear starting from a single active cell. 

Figures 6(a) and 6(b) show two snapshots of cellular structures at $t=5$ and 150 for $v_0=1.2,l_0=2$, and $L=42$. The membrane region with a small diffusion constant $D_0\sim 0.01D_1$ is marked in Fig.~6(a). Nine cells are created at $t=5$, and the cell number increases with time, as shown in Fig.~6(b). Figure 6(c) shows the time evolution of $D(x,y)$ at the section $y=L/2$. The cell number increases monotonically. 
Similar self-replicating structures were also studied in several models of artificial life.\cite{rf:14}
Figures 7(a) and 7(b) show two snapshots of the cellular state at $t=5$ and 75 for a smaller nutrient concentration $v_0=0.7$ and $l_0=2.5$. The system size is $L=30$.  Figure 7(c) displays the time evolution of $D(x,y)$ at the section $y=L/2$. Because the initially activated region is slightly larger than that in the case of Fig.~6, several cells are created at $t=5$. However, the cell number decreases with time and finally the cell structure disappears, which leads to a death state: $u=w=0$ for $v_0=0.7$. The extinction is caused by the lack of the nutrient V at the outer boundary. 

\section{Summary}
The membrane works as the boundary from the outside world, which makes a cell an independent element separated from the outside world. The membrane as a container is constructed by the contents of the cell, and a high activity of the contents of the cell is maintained and protected by the membrane.  We have constructed a stochastic Schl\"ogl model to understand the creation of the mutually dependent relation. We have shown with a Monte-Carlo simulation of the model that an active state surrounded by a semipermeable membrane is self-organized via a stochastic jump process starting from a death state in a homogeneous system with respect to diffusivity.  It can be interpreted as the birth of a cell. 
We have also constructed a breeding model of cell structures.  If the nutrient is sufficiently large, the cell number increases with the creation of the membrane, which leads to a multicellular system.  If the nutrient is not sufficient, the cell structure is extinguished. Our model is a toy model, but might be a suggestive model for considering a primitive cell or life from the view point of dissipative structures far from equilibrium.

\end{document}